# Thermo-mechanical characterization of on-chip buckled dome Fabry-Perot microcavities

M. H. Bitarafan,[1] H. Ramp,[2] T. W. Allen,[1] C. Potts,[1] X. Rojas,[2] A. J. R. MacDonald,[2] J. P. Davis,[2] and R. G. DeCorby[1*]

[1]ECE Department, University of Alberta, 2nd Floor, 9107-116 St. NW, Edmonton, AB, Canada, T6G 2V4
[2]Department of Physics, University of Alberta, 11322-89 Ave., Edmonton, AB, Canada, T6G 2G7
*Corresponding author: rdecorby@ualberta.ca

**Abstract:** We report on the thermo-mechanical and thermal tuning properties of curved-mirror Fabry-Perot resonators, fabricated by the guided assembly of circular delamination buckles within a multilayer a-Si/SiO$_2$ stack. Analytical models for temperature dependence, effective spring constants, and mechanical mode frequencies are described and shown to be in good agreement with experimental results. The cavities exhibit mode volumes as small as $\sim 10\lambda^3$, reflectance-limited finesse $\sim 3\times10^3$, and mechanical resonance frequencies in the MHz range. Monolithic cavity arrays of this type might be of interest for applications in sensing, cavity quantum electrodynamics, and optomechanics.

*OCIS codes:* (120.2230) Fabry-Perot; (130.3120) Integrated optics devices; (230.5750) Resonators.

## 1. Introduction

On-chip, high-finesse Fabry-Perot (FP) cavity arrays are of interest for lab-on-a-chip [1] and optomechanical [2] sensing systems. Compelling applications can also be found in the field of cavity quantum electrodynamics (CQED), where a major topic is the strong coupling between atoms and photons in an optical resonant cavity [3]. Optical cavities could potentially be the nodes within a 'quantum internet' [4-5], with information carried by single photons whose quantum state is manipulated at the nodes by interactions with atoms [6-7].

Although there are alternatives [3], the FP cavity is the prototypical structure for CQED [4-7]. To facilitate strong coupling (i.e. coherent interactions) between light and matter, the cavity should satisfy several key requirements [4,8-11]: (i) it should provide access to an air (or vacuum) core, so that atoms can be placed and trapped in the region of high photon density, (ii) it should have high finesse ($F$) and quality factor ($Q$), so that the decay rate of the cavity mode is small, (iii) it should have a small optical mode waist and volume, so that the atom-photon energy exchange rate is high, and (iv) it should be tunable so that the cavity can be brought into resonance with the atomic emitter. In addition, cavities should be sufficiently robust to survive and operate at low temperatures and in vacuum, and (where applicable) should exhibit high mechanical resonance frequencies [11]. It is anticipated that a quantum network will require arrays of tunable microcavities on a single chip [4,11-12].

Macroscopic curved mirror cavities with $F > 10^5$ but relatively large mode volume were reported more than 10 years ago [13]. Micro-machining techniques such as CO$_2$ laser ablation [8], focused-ion-beam (FIB) milling [10], and dry etching [14] have been studied in an effort to reduce size and enhance scalability. Often, one or both mirrors are formed on the end of an optical fiber [8], which provides a convenient means for light coupling. $F \sim 10^5$ and mode volumes as small as $\sim 40$ μm$^3$ have been achieved [9]. However, serial manufacturing approaches inhibit scalability, and fully monolithic integration strategies remain elusive [15-16]. Efforts towards the construction of high-finesse Fabry-Perot cavity *arrays* on a chip [10,15], particularly with individually tunable cavities [11], are at an early stage.

In a recent paper [17], we described curved-mirror, FP microcavities fabricated using a MEMS-like, thin film buckling technique. With this approach, the roughness of the mirror surfaces is determined mainly by deposition processes, rather than by a micro-machining process. Moreover, owing to their stress-driven self-assembly, the cavities exhibit an uncommon degree of morphological and optical predictability, including reflectance-limited finesse and textbook manifestations of Laguerre-Gaussian and Hermite-Gaussian modes. The technique enables straightforward fabrication of on-chip arrays, and the cavity size can be varied (within limits) through lithographic feature control. As shown below, a fundamental mode volume as small as $\sim 10\lambda^3$ has been realized. Since the buckled mirror is essentially a flexible plate, the cavities can be mechanically tuned and have potential for use in the study of optomechanics [2].

Understanding the thermal and mechanical properties [18] of the buckled microcavities is a prerequisite for the applications mentioned above. In the following, we describe the thermal dependence of the cavity resonance, which can be attributed primarily to the coupling between in-plane stress and out-of-plane deflection of the buckled mirror. We also describe the vibrational characteristics of the buckled mirrors, including mechanical resonance frequencies and effective spring constants. Approximate analytical theories are shown to be in good agreement with experimental observations.

## 2. Morphology of the buckled cavities

The buckled microcavities are essentially half-symmetric Fabry-Perot resonators (see Fig. 1), and their fabrication and optical properties were described previously [17]. Within a certain range of base diameters ($2a$), the profile of the buckled mirror is well approximated as a spherical dome segment (i.e. a shallow spherical shell). However, the exact shape is determined primarily by elastic buckling mechanics, influenced by secondary factors such as plastic deformation and relaxation of compressive stress over time. Assuming purely elastic deformation and perfectly clamped boundary conditions, the fundamental (axisymmetric) buckling profile for a circular delamination buckle can be expressed [19]:

$$\Delta(r) \approx \delta \cdot [0.2871 + 0.7129 \cdot J_0(\mu r)] \quad , \quad (1)$$

where $\Delta$ is the vertical deflection, $r$ is the radial coordinate (normalized to $a$), $\delta$ is the peak height of the buckle (see Section 3 below), $J_0$ is the Bessel function of first kind and order zero, and $\mu = 3.8317$.

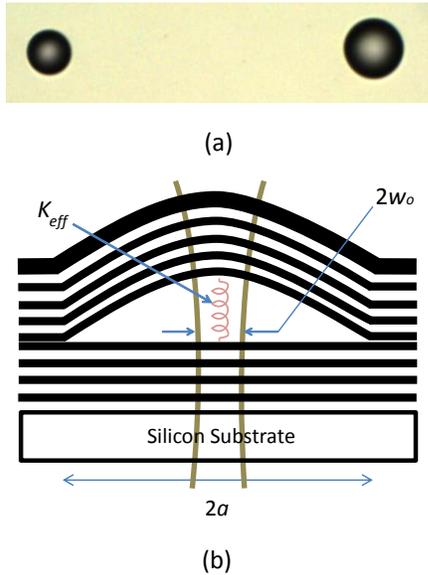

(a)

(b)

Fig. 1. (a) A microscope image showing two adjacent domes, one with 150 μm diameter and the other with 200 μm diameter. (b) Schematic illustration of a buckled dome microcavity in cross-section. The waist diameter of the fundamental optical mode ($2w_0$) is typically much less than the diameter of the dome base ($2a$). The upper buckled mirror is a flexible plate with quasi-clamped boundaries, and its movement is subject to an effective spring constant $K_{eff}$.

Figures 2(a) and 2(b) show the experimental cross-sectional profiles of typical 100 and 200 μm diameter cavities, compared to the shapes predicted for a dome and for a clamped, elastic buckle. The experimental profiles were obtained using an optical profilometer (Zygo NewView 5000). The dome and buckle models were normalized to the experimentally determined peak height in each case. For the dome model, curves are shown for two curvatures: $R_T$ is the curvature estimated from a fit near the top of the buckled mirror [17], while $R_D$ is the curvature for a dome that spans the same base diameter as the actual buckle. As illustrated by the data shown, the profile of smaller cavities is closer to the predictions of the elastic buckling theory while the profile of larger cavities is more dome-like. Generally speaking, the experimental profiles are intermediate with respect to the dome and buckle models.

Deviation from elastic behavior is not unusual for thin film delamination buckles [20]; plastic deformation near the boundaries can occur, and the assumption of clamped boundaries is often too simplistic. Nevertheless, using the measured pre-buckling compressive stress for the multilayer mirrors ($\sigma \sim 180$ MPa) in the elastic buckling model (see Eq. (3) below), good agreement between predicted and measured peak buckle heights was verified (see Fig. 2(c)).

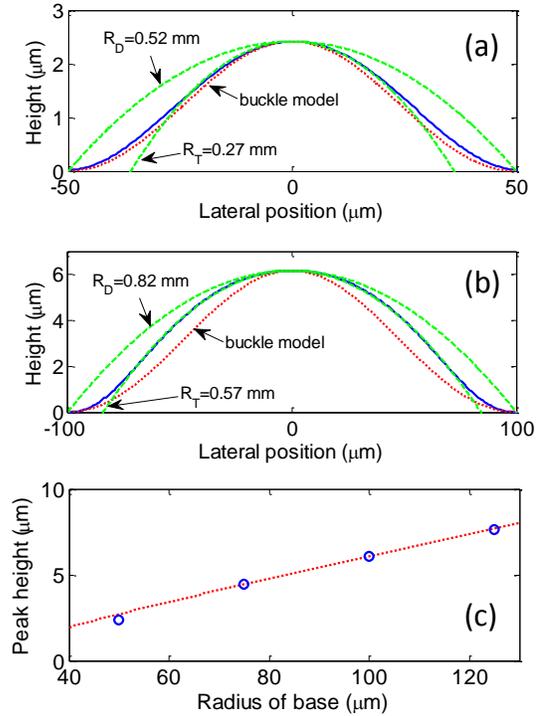

Fig. 2. (a) Experimental cross-sectional profile (blue solid line) for a typical 100 μm diameter cavity is compared to predictions based on a spherical dome assumption (green dashed lines) and a clamped circular buckle assumption (red dotted line). For the dome model, curves are shown for two different radii of curvature, as explained in the main text. (b) As in part (a), except for a typical 200 μm diameter cavity. (c) A plot of the peak buckle height versus base radius is shown. The red curve is the prediction of the elastic buckling model, assuming pre-buckling compressive stress of 180 MPa and the effective medium parameters shown in Table 1. The blue symbols are average values measured for cavities of varying base radius.

Given the complex shape of the cavities, an exact treatment of their mechanical properties would require numerical simulations. Here, we aim instead to estimate the main parameters of interest (mechanical resonance frequencies, spring constants, etc.) by employing analytical approximations. This provides significant insight while not obscuring the essential features. We invoke results from the literature on both shallow spherical shells (the terms 'shell' and 'dome' are used interchangeably in the following) and buckled circular plates. Encouragingly, predictions from both models are in good mutual agreement, and also show good agreement with experimental observations.

In keeping with an approximate approach, we treat the buckled mirror as a single plate characterized by effective-medium parameters (see Table 1). The mirror is a 4-period quarter-wave stack (QWS) with a half-wave amorphous Si (a-Si) capping layer, deposited by magnetron sputtering [17]. It has total thickness $h \sim 1.6$ μm and is ~37% a-Si and ~63% $SiO_2$ by volume. As is well known, thin films show significant variation in their thermal and elastic properties depending on deposition details. This is particularly the case for Young's modulus and the coefficient of thermal expansion (CTE), both of which play central roles in the analyses below. For these quantities, we based the effective medium parameters on values reported in the literature for similar a-Si (e.g. $E \sim 80$ GPa [20], $\alpha \sim 4.5 \times 10^{-6}$ [21]) and SiO$_2$ (e.g. $E \sim 60$ GPa, $\alpha \sim 3.1 \times 10^{-6}$ [22]) thin films. The other parameters in Table 1 were estimated from widely reported [23,24] values for SiO$_2$ and amorphous or polycrystalline Si thin films.

Table 1. Effective medium parameters assumed for the buckled mirrors.

| | Thickness | Density | Young's modulus | Poisson's ratio | Thermal expansion coefficient |
|---|---|---|---|---|---|
| Symbol | $h$ (μm) | $\rho$ (kg m$^{-3}$) | $E$ (GPa) | $\nu$ | $\alpha$ (K$^{-1}$) |
| Value | 1.6 | 2240 | 70 | 0.2 | 3.6x10$^{-6}$ |

## 3. Optical and thermal tuning properties

In a previous study [17], the optical properties of cavities with base diameters in the 200 to 400 μm range were reported. For the applications discussed above, cavities with even smaller dimensions (and mode volumes) are desirable. Consider the 100 μm diameter domes, which have peak height $\delta \sim 2.4$ μm and radius of curvature $R_T \sim 270$ μm. In the paraxial approximation, the beam waist (radius) for the fundamental mode of the half-symmetric cavity can be approximated as [8]:

$$w_0 \approx \sqrt{\lambda/\pi}(L \cdot R)^{1/4} \quad , \qquad (2)$$

where $L$ is the effective cavity length, $R$ is the radius of curvature for the curved mirror, and $L \ll R$ was assumed. Here, $L = \delta + 2d_P$, where $d_P$ is the phase penetration depth into the dielectric mirrors [25]. For operation near the stop-band center wavelength ($\lambda \sim 1.55$ μm here), $L \sim \delta + (\lambda/2)\{1/(n_H - n_L)\}$ [13], where $n_H$ and $n_L$ are the refractive indices of the high and low index layers. Using $n_H = 3.6$ and $n_L = 1.5$ gives $d_P \sim 200$ nm and $L \sim 2.8$ μm. Due to their high index contrast, the phase penetration depth is relatively small for these mirrors. Using $R = R_T$ (since the mode is confined to the central portion of the curved mirror), Eq. (2) then produces $w_0 \sim 3.7$ μm.

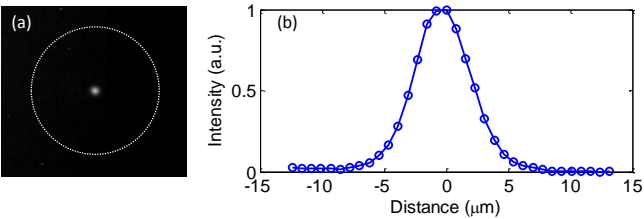

Fig. 3. (a) An image of the fundamental mode for a 100 μm diameter cavity is shown. The white dotted line indicates the dome boundary. (b) A plot of the transverse intensity profile for the fundamental mode from part (a) is shown. The 1/$e^2$ mode waist radius is ~4.5 μm.

To experimentally assess the mode size, a tunable laser (Santec TSL-320) was coupled to the cavity using a tapered lensed fiber (Oz Optics) with nominal focal spot diameter ~10 μm. The laser was tuned to the frequency of a fundamental resonance, in order to isolate and image the TEM$_{00}$ mode of the cavity [17]. Fig. 3(a) shows the mode field image captured using an infrared camera, and Fig. 3(b) shows a transverse intensity profile extracted from such an image. From the 1/$e^2$ intensity points, an experimental mode waist $w_0 \sim 4.5$ μm was estimated. This is in good agreement with the prediction above, especially given the limited pixel resolution of the camera images. For the standing-wave field associated with the TEM$_{00}$ mode, the effective mode volume can be approximated as $V_0 \sim (\pi/4) w_0^2 L$ [8]. For the 100 μm diameter cavity $V_0 \sim 10\lambda^3$; similar wavelength-scaled values have been reported for visible-band cavities [3,8-10,16].

The optical linewidth was studied using the tunable laser and a calibrated photodetector. It is worth noting that laser power was set low ($\ll 100$ μW) for all measurements described here, to avoid significant heating of the mirrors by laser absorption. At higher powers, we observed clear signatures of photo-thermal bistability and hysteresis [26]. Fig. 4(a) shows a typical fundamental resonance line for a 100 μm diameter cavity, with an input laser power of $\sim 3$ μW. The experimental linewidth (~0.16 nm) corresponds to $Q \sim 9600$ and finesse $F \sim Q/m \sim 3200$, where $m = 3$ is the longitudinal mode order for the cavity. This is in excellent agreement with the reflectance-limited finesse we reported for larger cavities with the same mirrors [17].

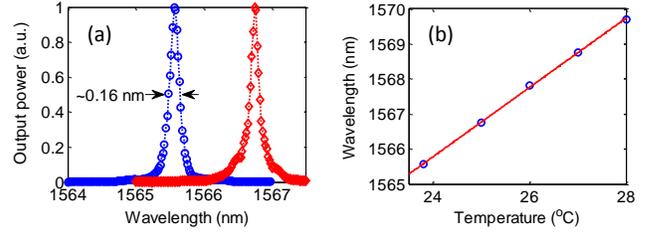

Fig. 4. (a) Experimental linewidth plots are shown for the fundamental resonance of a 100 μm dome, at 23.8 °C (circles) and 25 °C (diamonds). (b) The plot shows the variation in the fundamental resonance wavelength with temperature, revealing a red-shift $\Delta\lambda/\Delta T \sim 1$ nm/K. Blue symbols are experimental data points and red line is a linear fit to the data.

A unique feature of buckled structures is that in-plane stress is directly coupled with out-of-plane deflection [24,27-28]. We have previously developed and experimentally verified a model for the thermal tuning of straight-sided (Euler) delamination buckles [28], where out-of-plane deflection is driven by the difference in CTE between the buckled feature and the substrate. For a clamped *circular* plate, the critical buckling stress is given by [19] $\sigma_C = 1.2235[E/(1-\nu^2)](h/a)^2$, where $E$ is Young's modulus, $\nu$ is Poisson's ratio, and $h$ and $a$ are the thickness and radius of the plate. When compressive stress exceeds $\sigma_C$ (within limits), the plate buckles with an axisymmetric profile (see Eq. (1)) and peak height:

$$\delta = h\left[1.96\left(\sigma/\sigma_C\right) - 1\right]^{1/2} \quad , \qquad (3)$$

where $\sigma$ is the biaxial compressive stress and $\nu = 0.2$ was assumed. For a pre-existing circular buckle, an analogous treatment to that found in [28] leads to an estimate of the change in peak height with temperature:

$$\frac{\Delta\delta}{\Delta T} \approx 0.80(1+\nu)\frac{a^2}{\delta}\Delta\alpha \quad , \qquad (4)$$

where $\Delta\alpha$ is the difference in CTE between the buckled plate (i.e. the mirror) and the silicon substrate ($\alpha_{Si} \sim 2.5 \times 10^{-6}$) and $\delta$ is the initial peak height. For example, using $\Delta\alpha \sim 1.1\times10^{-6}$, Eq. (4) predicts $\Delta\delta/\Delta T \sim 1.1$ nm/K for $a = 50$ μm and $\delta = 2.4$ μm. Moreover, using $\Delta\lambda/\Delta T \sim (\lambda/\delta)(\Delta\delta/\Delta T)$, it follows that $\Delta\lambda/\Delta T \sim 0.7$ nm/K is predicted for the 100 μm diameter domes.

To corroborate this theory, samples were mounted on a thermo-electric cooler and scanned at various temperatures using either the optical profilometer to determine height changes or the tunable laser to determine changes in the spectrum. While the two types of measurements were in good general agreement, the spectral scans were more consistent and repeatable. As shown in Fig. 4(b) for a typical 100 μm diameter cavity, a red-shift of the resonant wavelength ($\Delta\lambda/\Delta T \sim 1$ nm/K) was observed, in good agreement with the theoretical prediction. Uncertainty in the CTE of the buckled mirror is probably the main source of residual discrepancy. Similar levels of agreement were found for the other cavity sizes.

The temperature dependence provides a convenient tuning mechanism, and integrated heater electrodes might even be feasible. However, this dependence could also be detrimental in some cases. For example, some CQED applications require resonance wavelength stability on the order of 1 pm [11], implying the need for a rather challenging temperature stability of ~ 0.001 K for the present cavities. This could be mitigated by matching the CTE of the mirror and substrate [23], and using an alternative (e.g. electrostatic) tuning mechanism.

## 4. Mechanical and dynamic properties

In order to exploit the buckled microcavities as sensors or optomechanical elements, a basic understanding of their mechanical and dynamical properties is required. This can be accomplished by employing a 'thermo-mechanical calibration' technique [18], where the random motion of a structure is extracted from the noise of a nominally steady-state signal. Here, the steady-state signal is the cavity transmittance at fixed laser detuning [16], and measurements of the noise on this signal yield the mechanical resonance frequencies $\omega_n$ for the upper (deformable) mirror. Within the limits of a classic harmonic oscillator model applied to each mode, these frequencies are related to the effective spring constant and mass of the mirror as $\omega_n = (K_{eff,n}/m_{eff,n})^{1/2}$. Furthermore, the mean-square amplitude of the fluctuations in mirror position (for a given mode) can be estimated by invoking the equipartition-of-energy theorem [18,26]:

$$\left\langle a_n^2(t) \right\rangle = \frac{k_B T}{m_{eff,n}\omega_n^2} = \frac{k_B T}{K_{eff,n}} \quad , \qquad (5)$$

where $k_B$ is the Boltzmann constant. In the following, we describe analytical approximations for the resonance frequencies and effective spring constants of the buckled mirror. These are corroborated by experimental results.

### 4.1. Vibrational resonance frequencies

As mentioned, the buckled mirror is analogous to a shallow spherical shell [27], so that analytical treatments from the theories of shells and plates are useful. The natural vibrational frequencies of a thin, flat, and clamped circular plate are well known [29], and can be expressed as $\omega_{P,n} = \Omega_{P,n}(1/a^2)(D/(\rho h))^{1/2}$, where $D = Eh^3/(12(1-\nu^2))$ is the flexural rigidity of the plate and tabulated values of $\Omega_{P,n}$ are available (e.g. $\Omega_{P,1} = 10.216$, $\Omega_{P,2} = 21.261$, $\Omega_{P,3} = 34.877$, etc. [29]). Soedel [30] showed that the natural frequencies for a shallow shell (i.e. a dome) can be estimated from those of the equivalent plate with the same projected boundary dimensions:

$$\omega_{S,n} = \sqrt{\omega_{P,n}^2 + E/(\rho R_S^2)} \quad , \qquad (6)$$

where $R_S$ is the shell radius of curvature. Consider for example the 200 μm diameter domes, and let $R_S \sim R_T = 0.57$ mm, justified by the excellent fit to the dome model in that case (see Fig. 2(b)). Using the effective medium parameters from Table 1, these equations predict $f_{P,1} = 430$ kHz and $f_{S,1} = 1.6$ MHz. We found that Eq. (6) provides accurate predictions of the lowest-order mechanical resonance frequency (especially for the larger cavities, as evidenced below), but is less accurate for the higher-order modes. This might be due to the fact that the fundamental (axi-symmetric) vibrational mode is most closely aligned with the central, spherical portion of the buckle. Furthermore, the shell formula neglects residual stress in the buckled plate [12].

An alternative approach is derived from the literature on the vibration of buckled structures [24,31]. For a symmetrically buckled structure, the resonance frequency of the lowest-order (i.e. symmetric) vibrational mode can be estimated as [32]:

$$\omega_{B,1} = \omega_{P,1} \cdot \sqrt{2} \cdot \sqrt{(\sigma/\sigma_C)-1} \quad , \qquad (7)$$

where $\sigma$ is the pre-buckling biaxial stress, $\sigma_C$ is the critical buckling stress (see Eq. (3)), and $\omega_{P,1}$ is the fundamental resonance frequency for the stress-free and flat plate from above. Encouragingly, Eq. (7) produces good agreement with the numerical results for a buckled circular plate reported by Williams et al. [31]. For the 200 μm diameter buckle $\sigma_C \sim 22.8$ MPa and (using $\sigma \sim 180$ MPa as above) Eq. (7) predicts $f_{B,1} = 1.6$ MHz, in good agreement with both the shell-based prediction and the experimental observations below.

As discussed above, experimental resonance frequencies can be obtained by observing the random thermo-mechanical motion of the buckled domes. We used a "tuned-to-slope" technique, similar to that used in other studies [16,37]. The frequency of a tunable laser was slightly detuned from an optical resonance, nominally at the point of maximum slope of the transmission. Random thermal motion of the buckled mirror changes the cavity length, shifts the optical resonance frequency, and hence changes the transmission through the structure. The time-dependent transmittance of the dome was captured and digitized using a high-speed analog to digital converter. This data was subsequently Fourier transformed and averaged to increase the signal to noise ratio.

Figure 5(a) shows a typical data set extracted from a 100 μm diameter dome in air and at room temperature. The set of peaks was fit to a series of Lorentz oscillator displacement spectral densities, using thermomechanical calibration techniques described elsewhere [18]. For example, the fit of the fundamental resonance line is shown in the plot. Note that the mechanical $Q$ (e.g. $Q \sim 75$ for the lowest-order mode in the case shown) is undoubtedly affected by squeeze-film damping and viscous damping due to collisions with air molecules [26]. It would be interesting to perform similar measurements in vacuum, and possibly at low temperature, but this is left for future work.

As shown in Fig. 5(b), the positions of the lowest-order vibrational modes were generally in excellent agreement with the theoretical predictions. For the shell model (Eq. (6)), we used $R_S = R_T$ as the best estimate of the actual plate curvature. $R_T \sim$

0.27, 043, 0.57, and 0.75 mm was experimentally estimated for the 100, 150, 200, and 250 μm diameter cavities, respectively. Except for the smallest cavities, this resulted in very good agreement between Eq. (6) and experimental data. On the other hand, the buckle model (Eq. (7)) produced reasonable agreement with experimental observations for all cavities studied.

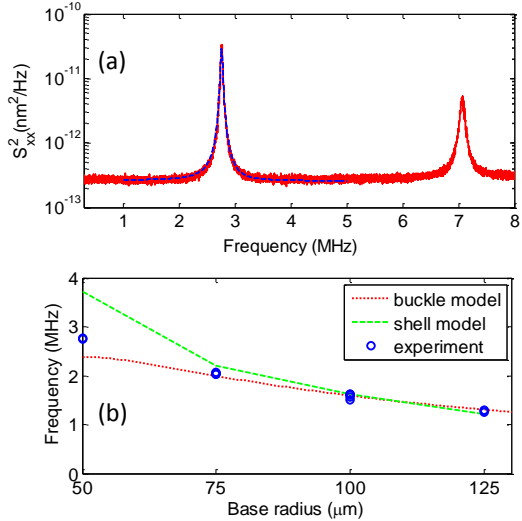

Fig. 5. (a) Mechanical spectrum arising from thermal noise, captured from a typical 100 μm diameter microcavity. $S^2_{XX}$ (red solid line) is the displacement spectral density of the measured time-domain signal [18]. The blue dashed line is a Lorentzian fit. (b) A plot of the fundamental mechanical resonance frequency versus radius of the cavity base is shown. The blue symbols are experimental data points; at least 3 of each size were measured, but data points overlap in some cases.

It is worth reiterating that these measurements were performed under ambient pressure, which introduces significant viscous damping of the mechanical motion. Moreover, relatively small-amplitude thermo-mechanical motion is predicted by Eq. (5) (given the relatively high effective spring constant of the present mirrors, discussed below). A numerical treatment of the buckled domes (not shown) confirmed that the low-order mechanical modes have similar wave-functions to those of the equivalent flat plate [29], implying that the optical and mechanical modes are characterized by a high degree of spatial overlap in these cavities. Thus, high optomechanical coupling coefficients are anticipated.

### 4.2. Effective stiffness (spring constants)

As per Eq. (5), thermo-mechanical calibration requires, in addition to the resonance frequencies, knowledge of the effective masses or spring constants [18]. The effective spring constant is also needed when assessing bistability and related effects [26]. It is important to note that $K_{eff,n}$ (as well as the effective mass of a given mode) will vary depending on how it is defined [18]. Moreover, deflection (and thus spring constant) will generally depend on the distributed nature of the load. Here, we will use the definition $K_{eff} = F/\Delta\delta$, where $\Delta\delta$ is the deflection of the buckle at its midpoint and $F$ is the applied force. Note that we have dropped the subscript '$n$', because the approximate theories presented below are not tailored or restricted to a specific vibrational mode.

Some forces of interest, such as the radiation pressure associated with the fundamental optical mode, are essentially concentrated (point) loads, while others, such as the photo-thermal force associated with changes in buckle temperature, are more closely approximated as distributed loads. Intuitively, we can expect a larger deflection (i.e. lower effective spring constant) if the force is concentrated near the center of the buckle. In the following, we discuss various approximations for the effective spring constant $K_{eff}$, and label them as $K_{I,J}$. Here the subscript $I$ refers to the use of a shell ($I=S$) or buckle ($I=B$) model, and the subscript $J$ refers to the assumption of a concentrated point ($J=P$) or uniformly distributed ($J=U$) load.

We first consider the concentrated load, and model the buckled mirror as a shallow shell. In this case, and for small deflections, an effective spring constant can be derived from the work by Lukasiewicz [33]:

$$K_{S,P} \approx \frac{\pi E h^2}{\sqrt{12(1-\nu^2)} \cdot R_S \cdot \left[\left(1/w^2\right)+\left(1/w\right)\text{ker}' w\right]} \quad , \quad (8)$$

where $w = w_1/l$, $w_1$ is the radius of the circularly symmetric, concentrated load applied to the center of the shell, and $l$ is a characteristic length for the shell:

$$l = \sqrt{R_S h}/\sqrt[4]{12(1-\nu^2)} \quad . \quad (9)$$

Furthermore, ker' is the first derivative of the Kelvin-real function [34]. Note that Eq. (8) does not contain the base radius $a$; this is because, for a load concentrated near the apex and for small deflections, the central deflection of the shell is approximately independent of the boundary conditions. Consider for example a 200 μm diameter cavity, and the case where $w_1 = w_0 \sim 5$ μm (i.e. the approximate size of the fundamental optical cavity mode [17]). This would describe the situation in which the mirror is deflected by radiation pressure forces. Using $R_S \sim R_T = 0.57$ mm (since the bending occurs primarily near the central part of the buckle in this case) and the other parameters from above, then $l \sim 16.4$ μm, $w \sim 0.3$, and $K_{S,P} \sim 800$ N m$^{-1}$.

Given the approximate nature of the shell analogy, it is useful to corroborate this result using the buckling literature. For a circular buckle, and in the limit of small deflections by a point load, an effective spring constant can be approximated from the numerical results of Jensen [35] (see Fig. 8 of that manuscript):

$$K_{B,P} \approx \frac{32\pi \cdot E \cdot h^3}{a^2 \cdot 3(1-\nu^2)} \quad , \quad (10)$$

which produces $K_{B,P} \sim 1000$ N m$^{-1}$ for the 200 μm cavity, in reasonable agreement with $K_{S,P}$.

Of greater interest here is the response to a distributed force (i.e. $F = P\pi a^2$, where $P$ is a uniform pressure) such as the thermal Langevin force that drives thermo-mechanical motion. As mentioned, a higher effective spring constant is anticipated in this case, and this is supported by results from the literature on delamination buckles. In the limit of small deflections, the deflection for a point-loaded circular buckle is four times that for a uniformly loaded buckle [19]. It follows that $K_{B,U} \sim 4K_{B,P} \sim 4000$ N m$^{-1}$ should be a reasonable approximation for the 200 μm cavity.

As above, we seek to corroborate this result by considering the literature on shallow spherical shells. From the work by Jones

[36], an effective spring constant for a uniformly loaded shell can be derived:

$$K_{S,U} \approx 64\pi \cdot D \left[1 + (1+\nu)\left(\frac{a^4}{8R_S^2 h^2}\right)\right] \bigg/ a^2 \quad . \quad (11)$$

It is somewhat problematic to define the radius of curvature for the real structures, as discussed in Section 2. However, it is reasonable to use $R_S \sim R_D$ in Eq. (11), because of the strong dependence on the dome radius $a$. For the 200 μm dome cavity with $R_D = 0.82$ mm, we find $K_{S,U} \sim 4800$ N m$^{-1}$, which is in good agreement with the buckle estimate. As shown in Table 2, Eq. (11) predicts that $K_{S,U}$ is fairly insensitive to the cavity size for the structures studied. These cavities are quite rigid in comparison to many MEMS-based cavities [11], and thus should suffer less from thermally induced degradation of the optical finesse [26].

Table 2. Estimated spring constants and effective masses

| Base diameter (μm) | $m_B$ (ng) | $R_D$ (mm) | $K_{S,U}$ (N/m) | $m_{eff,1}/m_B$ |
|---|---|---|---|---|
| 100 | 28 | 0.52 | 4.7x10$^3$ | 0.56 |
| 150 | 63 | 0.64 | 5.0x10$^3$ | 0.47 |
| 200 | 113 | 0.82 | 4.8x10$^3$ | 0.42 |
| 250 | 176 | 1.02 | 4.7x10$^3$ | 0.40 |

These estimates are expected to be correct to first-order only, especially since the shell and buckle models do not exactly describe the real structures. Nevertheless, the experimental data on the fundamental resonance frequencies suggests that the approximations are reasonable, as follows. Combining $\omega_1$ and $K_{S,U}$ produces an estimate for the effective mass of the fundamental vibrational mode: $m_{eff,1} \sim K_{S,U}/\omega_1^2$. For example, using the data from Fig. 5 and Table 2 produces $m_{eff,1} \sim 48$ ng for the 200 μm cavity. The buckled mirror has a total mass $m_B \sim 113$ ng in that case, and the ratio $m_{eff,1}/m_B \sim 0.42$ is quite reasonable for the fundamental vibrational mode of a circular plate [18]. As shown in Table 2, similarly reasonable results were obtained for the other cavity sizes. A more precise numerical analysis, and a more detailed experimental study of cavity stiffness, is left for future work.

## 5. Discussion and Conclusions

The buckled dome microcavities can be fabricated in large arrays, and might provide an interesting platform for sensing, CQED, and optomechanical coupling studies [1-3]. The mode volume and finesse demonstrated above are well within the ranges required to achieve strong coupling in CQED experiments [8,11]. Moreover, the finesse of the cavities might be increased by reducing the absorption loss in the mirrors, for example by using hydrogenated amorphous silicon for the high index layers. It should be noted that most CQED studies to date use Rb atoms and operate in the 700-800 nm wavelength range. In principle, it should be possible to fabricate compatible buckled microcavities using alternative mirrors based on SiO$_2$/TiO$_2$ or SiO$_2$/Ta$_2$O$_5$. The development of such a process, including control over adhesion and stress in these material systems, would be an interesting avenue for future study.

For many of the applications mentioned, it is necessary to incorporate 'open access' to the hollow cavity of the nominally enclosed buckle. It might be possible to incorporate this functionality directly into the buckling process by creating an on-chip network of intersecting hollow channels and microcavities [17]. However, a simpler alternative might be to machine 'micropores' or 'nanopores' directly through the upper mirror using a technique such as focused-ion-beam milling [38]. We hope to explore these options in future work.

### Acknowledgement

The authors gratefully acknowledge financial support from the Natural Sciences and Engineering Research Council of Canada (NSERC), Alberta Innovates Technology Futures (AITF), Canada Foundation for Innovation (CFI), and the Alfred P. Sloan Foundation.